\documentclass[reprint,superscriptaddress,amssymb,amsmath,aps,showpacs,10pt,floatfix,prb,longbibliography]{revtex4-2}
\def\YVS{YV$_6$Sn$_6$}
\def\doped{YV$_6$Sn$_{4.6}$Sb$_{1.4}$}
\def\cm{cm$^{-1}$}

\usepackage{graphicx}%
\usepackage{color}
\usepackage{epstopdf}
\usepackage{nicefrac}
\usepackage{lmodern}
\usepackage{amssymb}
\usepackage[T1]{fontenc}
\usepackage{amsmath}
\usepackage{amsfonts}
\usepackage[note-name=, use-sort-key = false]{notes2bib}

\usepackage{xcolor}

\usepackage{color}
\usepackage[colorlinks,bookmarks=false,citecolor=darkblue,linkcolor=red,urlcolor=blue]{hyperref} 
\definecolor{darkred}{rgb}{0.7,0.0,0.0}

\definecolor{darkblue}{rgb}{0,0.02,0.45}

\definecolor{darkgreen}{rgb}{0.02,0.45,0.0}

\definecolor{violet}{rgb}{0.8,0.2,0.6}

\begin{document}
\title{Rigid Band Shift in Electron-Doped Kagome Metal}

\author{M. Pinteri\'{c}}
\affiliation{1. Physikalisches Institut, Universit{\"a}t Stuttgart, 70569 
Stuttgart, Germany}
\affiliation{Faculty of Civil Engineering, Transportation Engineering and Architecture, University of Maribor, 2000 Maribor, Slovenia}

\author{E. Uykur}
\affiliation{Helmholtz-Zentrum Dresden-Rossendorf, Institute of Ion Beam Physics and Materials Research, 01328 Dresden, Germany}

\author{S. Salters}
\affiliation{ Materials Department, University of California Santa Barbara, Santa Barbara, California 93106, USA}

\author{G. Pokharel}
\affiliation{ Perry College of Mathematics, Computing, and Sciences University of West Georgia, Carrollton, Georgia 30118, USA}

\author{S. D. Wilson}
\affiliation{ Materials Department, University of California Santa Barbara, Santa Barbara, California 93106, USA}

\author{M. Dressel}
\affiliation{1. Physikalisches Institut, Universit{\"a}t Stuttgart, 70569 
Stuttgart, Germany}

\author{M. Wenzel}
\email{maxim.wenzel@pi1.physik.uni-stuttgart.de}
\affiliation{1. Physikalisches Institut, Universit{\"a}t Stuttgart, 70569 
Stuttgart, Germany}

\date{\today}

\begin{abstract}
We study the effects of Sn-site doping on the electronic structure of nonmagnetic kagome metal YV$_6$Sn$_{6-x}$Sb$_x$ by infrared spectroscopy and density-functional theory. The results for samples with $x$ = 0 and 1.4 demonstrate that partial substitution of Sn by Sb acts as a rigid shift of the Fermi level while leaving the band dispersion largely intact. Our findings establish Sn-site substitution in YV$_6$Sn$_6$ as a route to systematically investigate band-filling-dependent electronic properties of nonmagnetic kagome metals.
\end{abstract}

\maketitle

\section{Introduction}
Kagome metals are layered materials in which itinerant 3$d$ electrons occupying the metallic kagome lattice give rise to a characteristic electronic structure featuring flat bands, Van Hove singularities, and linearly dispersing bands, which can drive a variety of electronic orders~\cite{Denner2021, Liu2019}. The physical properties of these systems are highly sensitive to the band filling, which, together with electronic correlations and electron-phonon coupling, shapes their rich phase diagrams. The emerging phases include exotic magnetic orders, charge- and spin-density waves, unconventional superconductivity, and nematicity~\cite{Kiesel2013, Profe2024, Zhan2026, Wilson2024, DiSante2026, Jiang2022}.

$RM_6X_6$ (in short '166') kagome metals, with $R$ = rare-earth element, $M$ = Mn, Fe, V, Cr, Co, or Nb, and $X$ = Sn or Ge, provide a versatile platform for studying kagome physics, offering a broader chemical space than the widely studied $A$V$_3$Sb$_5$ family~\cite{Ortiz2025}. These compounds exhibit a variety of electronic and magnetic phenomena, including helical magnetic orders~\cite{Venturini1991, Ghimre2020} and charge-density-wave (CDW) order in ScV$_6$Sn$_6$~\cite{Arachchige2022} and LuNb$_6$Sn$_6$~\cite{Ortiz2025}. Their electronic structure features characteristic signatures of the kagome lattice, including Dirac-like crossings and multiple Van Hove singularities in close proximity to the Fermi level~\cite{Hu2024, Yi2024, Yang2024}. The broad chemical flexibility of the 166 family therefore provides an attractive setting for exploring how changes in band filling influence the electronic properties of kagome systems.

The CDW states observed in ScV$_6$Sn$_6$ and LuNb$_6$Sn$_6$ have transition temperatures comparable to those of the $A$V$_3$Sb$_5$ compounds. In contrast to $A$V$_3$Sb$_5$, however, these CDW states are associated with structural modifications involving predominantly out-of-plane displacements of the Sn atoms, while the kagome network and the V/Nb-derived electronic states remain largely unaffected~\cite{Arachchige2022, Korshunov2023, Hu2025, Lee2024, Lou2025}. Competing ordering vectors have been linked to the softening of flat phonon modes and the resulting structural instability~\cite{Wang2024, Korshunov2023, Hu2025, Pokharel2023, Yang2025, Alvarado2024, Cao2023, Tan2023}. This structural instability has been attributed to the size of the ion occupying the $R$ site~\cite{Meier2023}. In ScV$_6$Sn$_6$, the CDW can be suppressed by as little as 5\% substitution of Sc by Y or Lu and is generally absent in 166 compounds with sufficiently large $R$-site ions.

Here, we investigate the possibility of tuning the Fermi level in nonmagnetic YV$_6$Sn$_6$~\cite{Pokharel2021, Rosenberg2024} by replacing Sn. We perform Fourier-transform infrared spectroscopy and density-functional theory calculations and find that Sb substitution modifies the low-energy electrodynamic response while leaving the underlying band dispersion largely intact. The observed evolution can be understood in terms of a rigid shift of the Fermi level relative to the band structure. Our findings demonstrate that substitution on the $p$-block site provides a route to tuning the Fermi level in nonmagnetic 166 kagome metals, enabling to study how band filling influences their electronic properties in systems where the $R$-site-size-driven structural instability is suppressed.

\section{Methods}
\subsection{Experimental}
The samples were synthesized using a self-flux growth method. The powders were weighed and mixed, with a 30:1 excess of tin, in the glovebox into a Canfield crucible before being sealed into an ampule. The ampule was flame-sealed to preserve an inert atmosphere to prevent yttrium decomposition. It was then placed in a high-temperature box furnace, rapidly heated to 1125~°C, and allowed to slow cool (2~°C/hr) to 780~°C. The ampule was then removed from the furnace and centrifuged to remove the tin flux, and the crystals were extracted. Due to excess tin on the surface, the samples were polished to expose the pure crystal facet, and EDX was used to determine doping composition.  

Prior to optical measurements, four-probe dc resistivity curves were measured. Temperature-dependent reflectivity measurements from room temperature down to 10~K were performed in the isotropic $ab$-plane on optically polished samples with lateral dimensions of $\sim 1 \times 1$~mm$^2$ and a thickness of around 500~$\mu$m. A broad frequency range from 80 to 18000~\cm\ (10~meV -- 2.23~eV) was covered utilizing different Bruker Fourier-transform infrared spectrometers. Data in the high-energy range ($\omega\,>\,600\,\mathrm{cm^{-1}}$) were collected using a Bruker Vertex 80v spectrometer coupled with a Hyperion IR microscope and a CryoVac Konti cryostat, while the low-energy range ($\omega\,<\,600\,\mathrm{cm^{-1}}$) was measured with a Bruker IFS113v spectrometer and a custom-built cryostat. Freshly evaporated gold mirrors served as reference in these measurements. The absolute value of the reflectivity was obtained by an in-situ gold-overcoating technique in the far-infrared range.

Below 80~\cm, standard Hagen-Rubens extrapolations were used, considering the metallic nature of the samples, while at high energies, x-ray scattering functions were utilized to extrapolate the data~\cite{Tanner2015}. The optical conductivity was then calculated from the measured reflectivity via Kramers-Kronig analysis~\cite{Dressel2002}.

\subsection{Computational}
Density-functional-theory (DFT) calculations of the band structure and optical conductivity were performed in the \texttt{Wien2K}~\cite{wien2k, Blaha2020} code with the Perdew-Burke-Ernzerhof flavor of the exchange-correlation potential~\cite{pbe96}. Experimentally determined structural parameters published in Ref.~\cite{Pokharel2021} were used. Fully relativistic self-consistent calculations were converged on the $12\times 12\times 6$ $k$-mesh. The optical conductivity was calculated using the \texttt{OPTIC} module~\cite{Draxl2006} on a denser $k$-mesh with up to $57\times 57\times 30$ points. 

\section{Results and Discussion}

\subsection{Dc Transport}
\begin{figure}
	\centering
	\includegraphics[width=1\columnwidth]{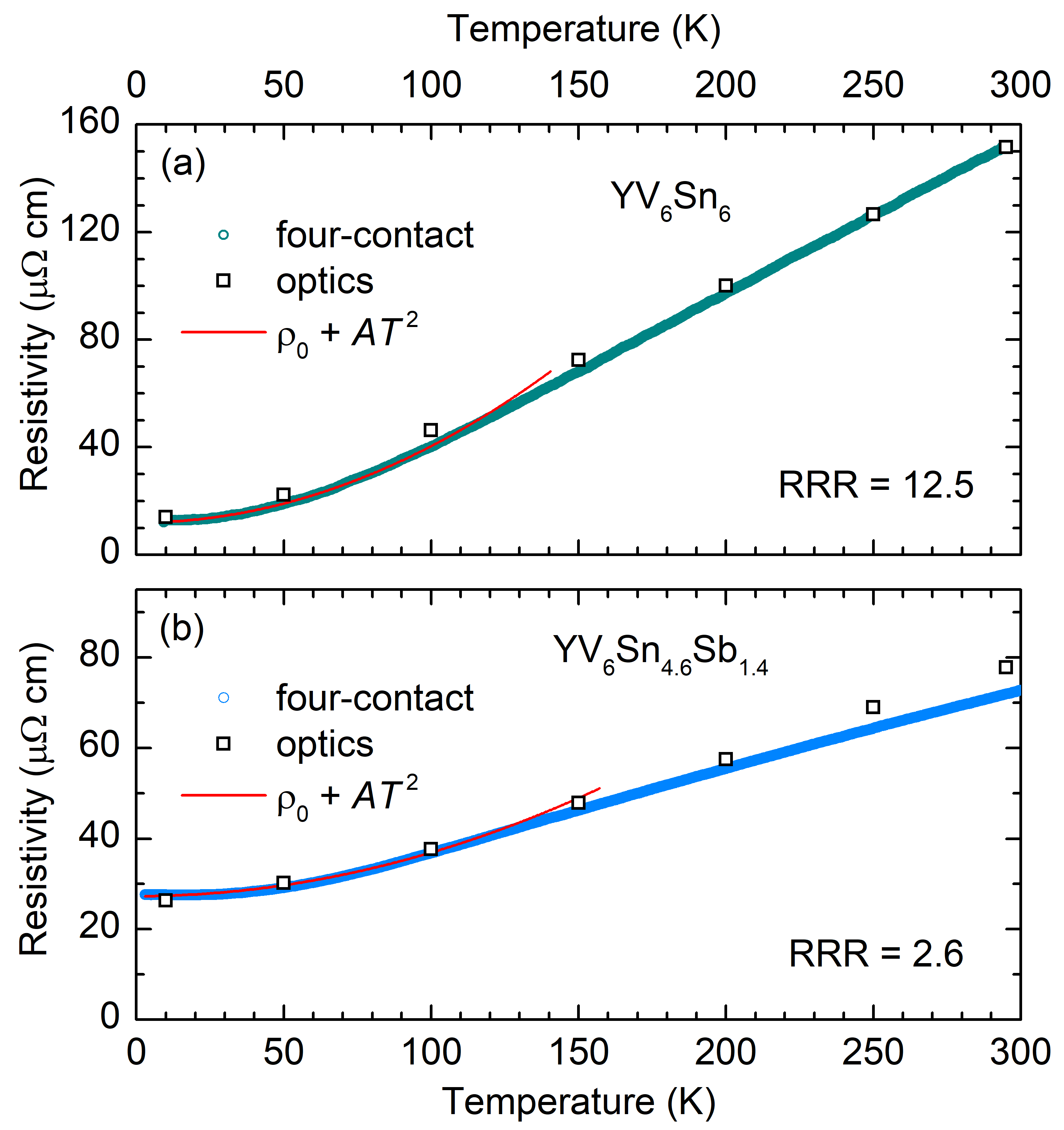}
	\caption{Four-probe dc resistivity of \YVS\ (a) and \doped\ (b) measured on the samples used in the optical experiments. Black squares represent the resistivity values obtained from the Hagen-Rubens fits to the low-energy reflectivities. }		
	\label{resistivity}
\end{figure}

The temperature-dependent in-plane dc resistivity of \YVS\ and \doped\ is presented in Fig.~\ref{resistivity}. The pristine \YVS\ sample exhibits a residual resistivity ratio (RRR) of 12.5, in good agreement with previous reports~\cite{Pokharel2021, Shtefiienko2025}. This ratio is significantly suppressed in the Sb-doped sample. As shown by the red fitting curve, the resistivity data are well described by $\rho = \rho_0 + AT^2$ up to $\sim$~125~K, consistent with electron-electron scattering. For pristine \YVS, we obtain $\rho_0 = 11.89\,\mu\Omega$cm and $A = 0.00285\,\mu\Omega$cm K$^{-2}$. In the doped sample, $\rho_0$ increases considerably to $27.04\,\mu\Omega$cm, while $A$ decreases to $0.000964\,\mu\Omega$cm K$^{-2}$. The significant decrease of the $A$ constant is in line with the increased carrier concentration and the larger Fermi surface of the Sb-doped sample, as discussed below.

Overall, the suppressed RRR and enhanced residual resistivity in \doped\ may be attributed to increased disorder introduced by the Sn-site substitution, while the compound remains highly metallic. The reduction in RRR, however, is not solely a consequence of the enhanced $\rho_0$ but is also influenced by the lower resistivity of the doped compound in the high-temperature, approximately linear-in-$T$ regime. In this regime, \doped\ exhibits a resistivity slope of $0.178\,\mu\Omega$/K, substantially lower than the $0.580\,\mu\Omega$/K observed for pristine \YVS.

\subsection{Optical Spectra}
\begin{figure}
	\centering
	\includegraphics[width=1\columnwidth]{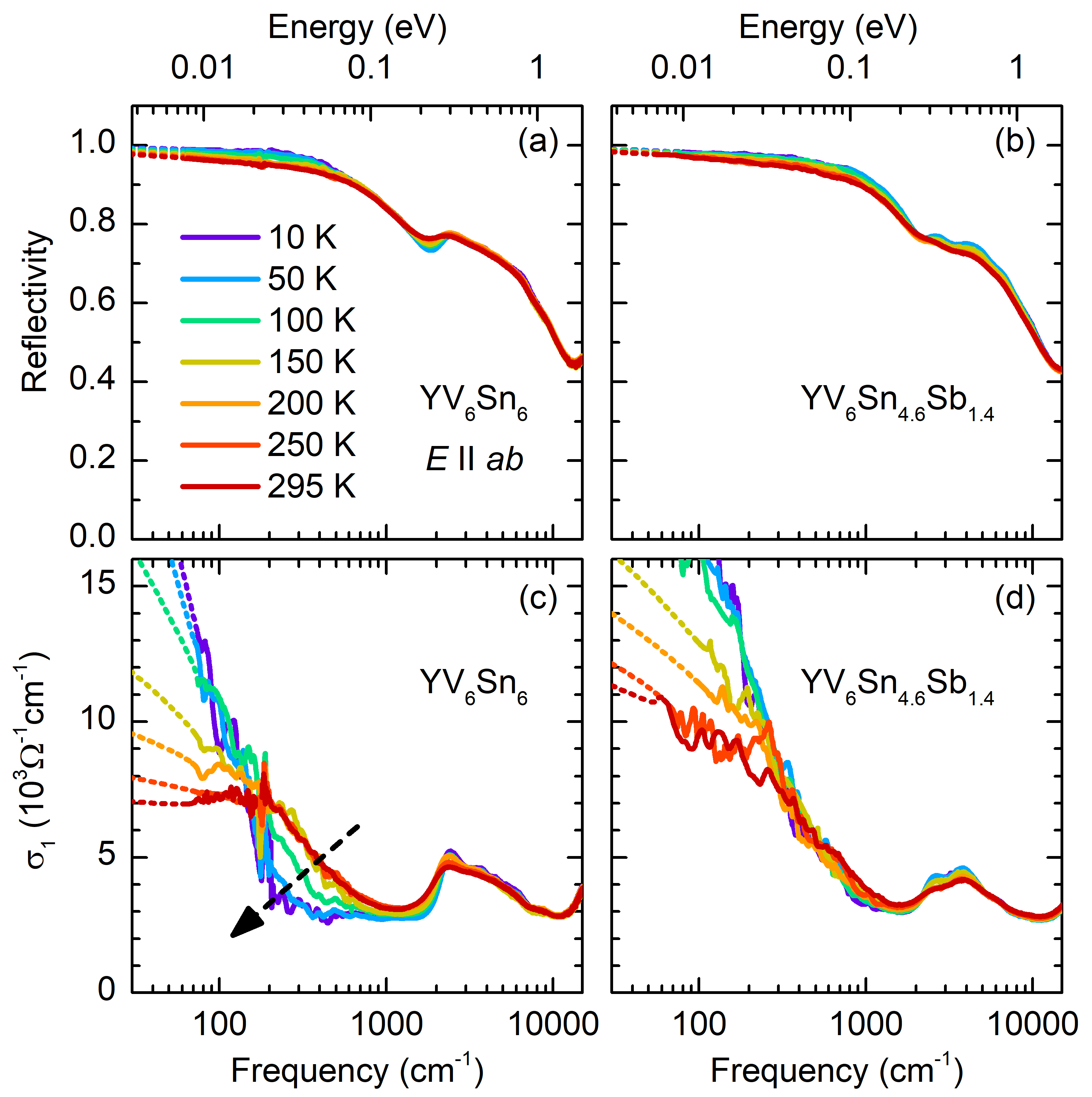}
	\caption{(a)(b) Temperature-dependent reflectivity of \YVS\ and \doped, respectively, measured in the $ab$ plane. (c)(d) Corresponding real parts of the optical conductivity. The low-energy values obtained by the Hagen-Rubens extrapolation are plotted with dotted lines. }		
	\label{ROC}
\end{figure}

Panels (a) and (b) of Fig.~\ref{ROC} show the reflectivity over a broad frequency range as a function of temperature for the undoped and doped \YVS\ samples, respectively. Consistent with the absence of anomalies in the resistivity curves, the continuous temperature evolution of the optical spectra confirms the absence of a structural or electronic phase transition in both the undoped and doped compounds.

\begin{figure}
	\centering
	\includegraphics[width=1\columnwidth]{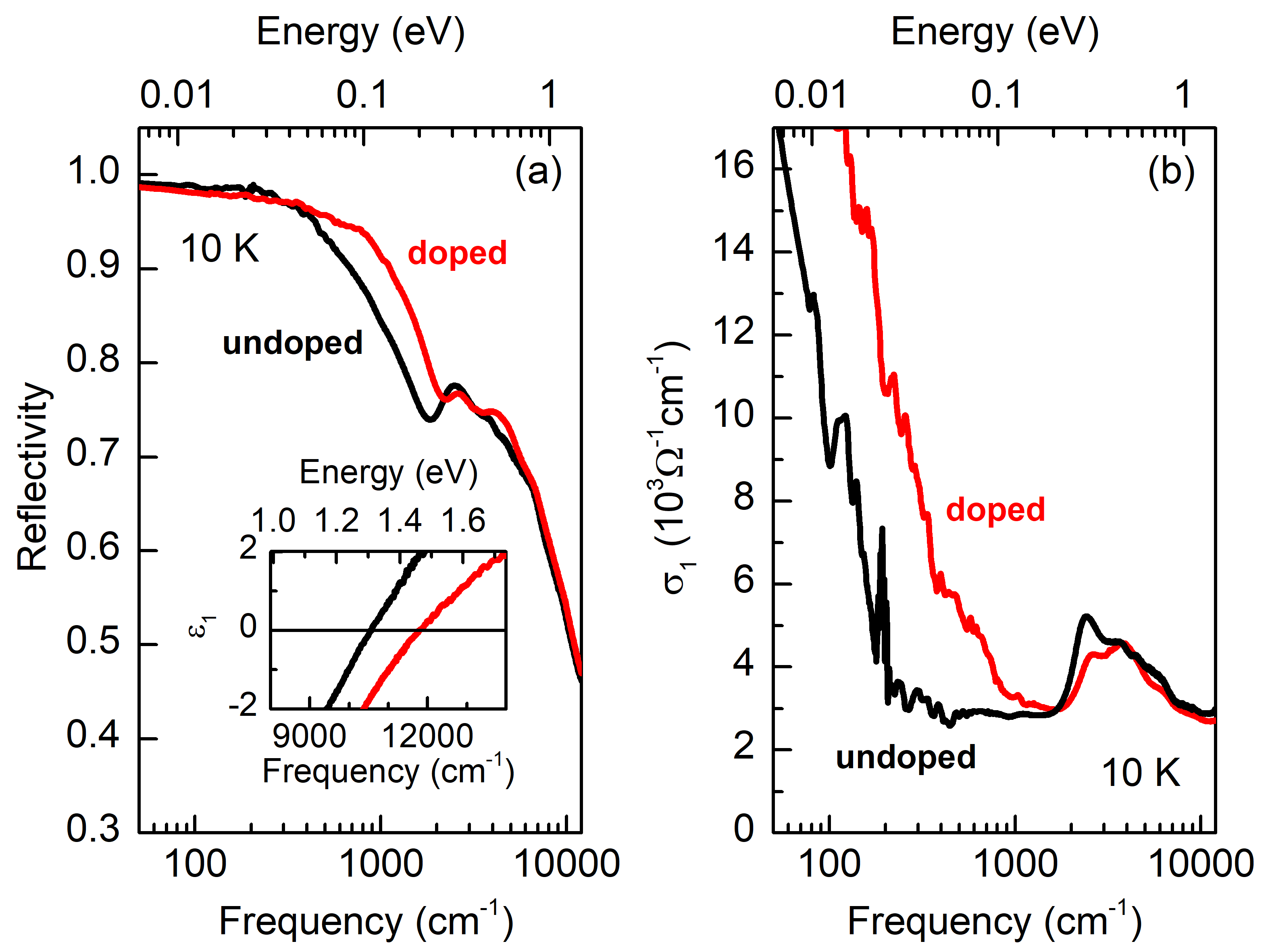}
	\caption{Comparison of the optical spectra of pristine \YVS\ and \doped\ at $T = 10$~K. The measured reflectivity is plotted in panel (a). The inset shows the zero crossing of the real part of the dielectric permittivity, marking the screened plasma frequency $\omega_p^{\mathrm{screened}} = \omega_p/\sqrt{\varepsilon_{\infty}}$. Panel (b) presents the calculated real part of the optical conductivity.}		
	\label{comparison}
\end{figure}

At first glance, both compounds display highly similar reflectivity curves with comparable temperature dependence. However, a closer comparison at 10~K, shown in Fig.~\ref{comparison}(a), reveals substantial changes in the spectrum of the doped sample. In particular, the plasma edge shifts to higher energies, also reflected in the zero crossing of the real part of the dielectric function (see inset), corresponding to the screened plasma frequency, $\omega_p^{\mathrm{screened}} = \omega_p/\sqrt{\epsilon_\infty}$. Here, $\epsilon_\infty$ represents the high-energy contributions to the real part of the dielectric function.

Additional changes in the optical spectra are recognized in the real part of the optical conductivity, plotted in Fig.~\ref{ROC}(c) and (d). Most prominently, the pronounced temperature dependence in the far-infrared range observed in pristine \YVS\ (marked by the black arrow) is absent in the doped sample, where the spectra remain nearly temperature-independent down to $\sim$~300~\cm. This contrasting behavior suggests that the low-energy optical response in pristine \YVS\ is predominantly associated with strongly temperature-dependent intraband transitions, which sharpen upon cooling as a result of the decreasing scattering rate, whereas the temperature-independent response in the doped sample may indicate significant contribution from interband transitions in this spectral range. Furthermore, the characteristic wiggle-like absorption features between 2000 and 6000~\cm\ are altered upon doping, as seen in Fig.~\ref{comparison}(b), indicating further modifications of the interband optical response in the mid-infrared range.

In addition, the spectra of pristine \YVS\ show a phonon mode centered around 180~\cm, which can be assigned to the in-plane $E_{1u}$ mode commonly observed in the optical spectra of 166-type kagome metals~\cite{Wenzel2022, Kim2023}. This mode could not be resolved in the optical spectra of the doped sample. Its absence, however, does not imply a structural modification or a change in phonon dynamics. Rather, the increased metallicity observed in the doped sample may lead to enhanced screening of the phonon by itinerant charge carriers.

\subsection{Decomposition}
To further separate and discuss modifications in the intra- and interband optical response, we model the optical spectra by the Drude-Lorentz approach. The complex dielectric function [$\tilde{\varepsilon}(\omega) = \varepsilon_1(\omega) + \mathrm{i}\varepsilon_2(\omega)$] is described by a Drude term accounting for the response of itinerant charge carriers and a sum of Lorentz oscillators describing interband transitions and the phonon mode:
\begin{equation}
\label{Eps}
\tilde{\varepsilon}(\omega)= \varepsilon_\infty - \frac{\omega^2_{p,{\rm Drude}}}{\omega^2 + i\omega/\tau_{\rm\, Drude}} + \sum\limits_j\frac{\Omega_j^2}{\omega_{0,j}^2 - \omega^2-i\omega\gamma_j}.
\end{equation}
Here, $\omega_{p,{\rm Drude}}$ and $1/\tau_{\rm\,Drude}$ are the plasma frequency and the scattering rate of the itinerant carriers, respectively. $\omega_{0,j}$, $\Omega_j$, and $\gamma_j$ describe the resonance frequency, strength, and the width of the $j^{th}$ excitation, respectively. The complex optical conductivity [$\tilde{\sigma}(\omega)=\sigma_1(\omega) + i\sigma_2(\omega)$] is then obtained by
\begin{equation}
\tilde{\sigma}(\omega)= -i\omega[\tilde{\varepsilon}(\omega)-\varepsilon_\infty]/4\pi.
\label{Cond}
\end{equation}

\begin{figure}
	\centering
	\includegraphics[width=1\columnwidth]{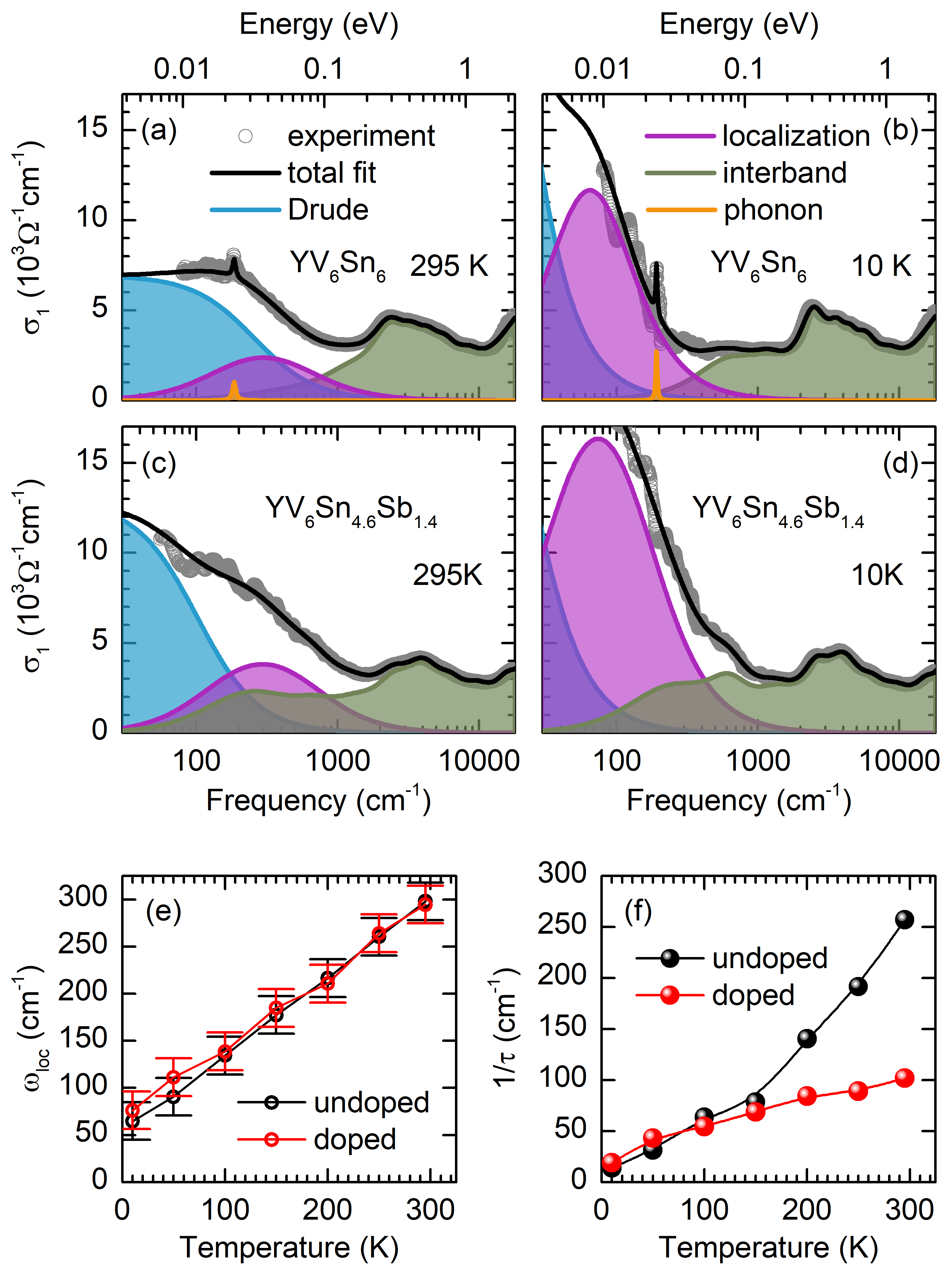}
	\caption{Exemplary decompositions of the optical conductivity of \YVS\ at room temperature (a) and 10~K~(b). (c)(d)~Same but for the doped sample. The spectra are modeled by a Drude peak (blue), a localization peak (purple), and multiple Lorentzians describing the interband transitions (dark green). In pristine \YVS, an additional phonon mode is resolved (orange). (e)~Temperature evolution of the position of the localization peak. Error bars of 20~\cm\ are assumed. (f)~Drude scattering rates obtained from the fits. }		
\label{decomp}
\end{figure}

Examples of the decomposed optical conductivity at $T = 295$~K and 10~K are presented in Fig.~\ref{decomp}. Lorentz oscillators are used to describe the interband transitions (dark green), which remain largely unchanged with temperature and only undergo sharpening upon cooling. In addition to these contributions, we identify a strongly temperature-dependent absorption feature that shifts to lower energies with decreasing temperature (purple) and therefore should not be interpreted as an interband feature. At the lowest temperatures, only the high-energy tail of this absorption remains visible the measured frequency range. In line with previous optical studies of kagome metals, we interpret this feature within the framework of a displaced Drude peak associated with strong electron-phonon interactions leading to localized carriers~\cite{Fratini2014, Rammal2024, Wenzel2022, Uykur2022, Wenzel2025, Wenzel2023}. 

We note that a previous optical study on \YVS\ described the intraband response using two Drude components with distinct scattering rates~\cite{Kim2023}. While this interpretation is well justified for \YVS, the observation of pronounced localization peaks in the optical spectra of several kagome materials, including the closely related GdMn$_6$Sn$_6$, supports a common description of the intraband absorption across various families of kagome metals~\cite{Wenzel2022, Uykur2022, Biswas2019}. The temperature dependence of the localization peak position exhibits a linear-in-$T$ redshift, as shown in Fig.~\ref{decomp}(e), closely resembling both the magnitude and temperature dependence of the scattering rate of the broad Drude component reported in Ref.~\cite{Kim2023}.

While the response of localized carriers is not significantly affected by doping, the low-energy upturn in the optical conductivity, modeled by a Drude contribution, is strongly modified by Sb substitution. The enhanced low-frequency optical conductivity of \doped\ is consistent with its lower resistivity. For the fits, we fix the dc conductivities to the values obtained from the Hagen-Rubens extrapolations. The scattering rates obtained from the modeled Drude contributions are shown in Fig.~\ref{decomp}(f) and closely follow the temperature dependence of the dc transport. In particular, pristine \YVS\ exhibits significantly higher scattering rates above $\sim$~150~K, consistent with the larger slope in the linear-in-$T$ regime of the dc resistivity (see Fig.~\ref{resistivity}).

\subsection{Interband Transitions}
\begin{figure}
	\centering
	\includegraphics[width=1\columnwidth]{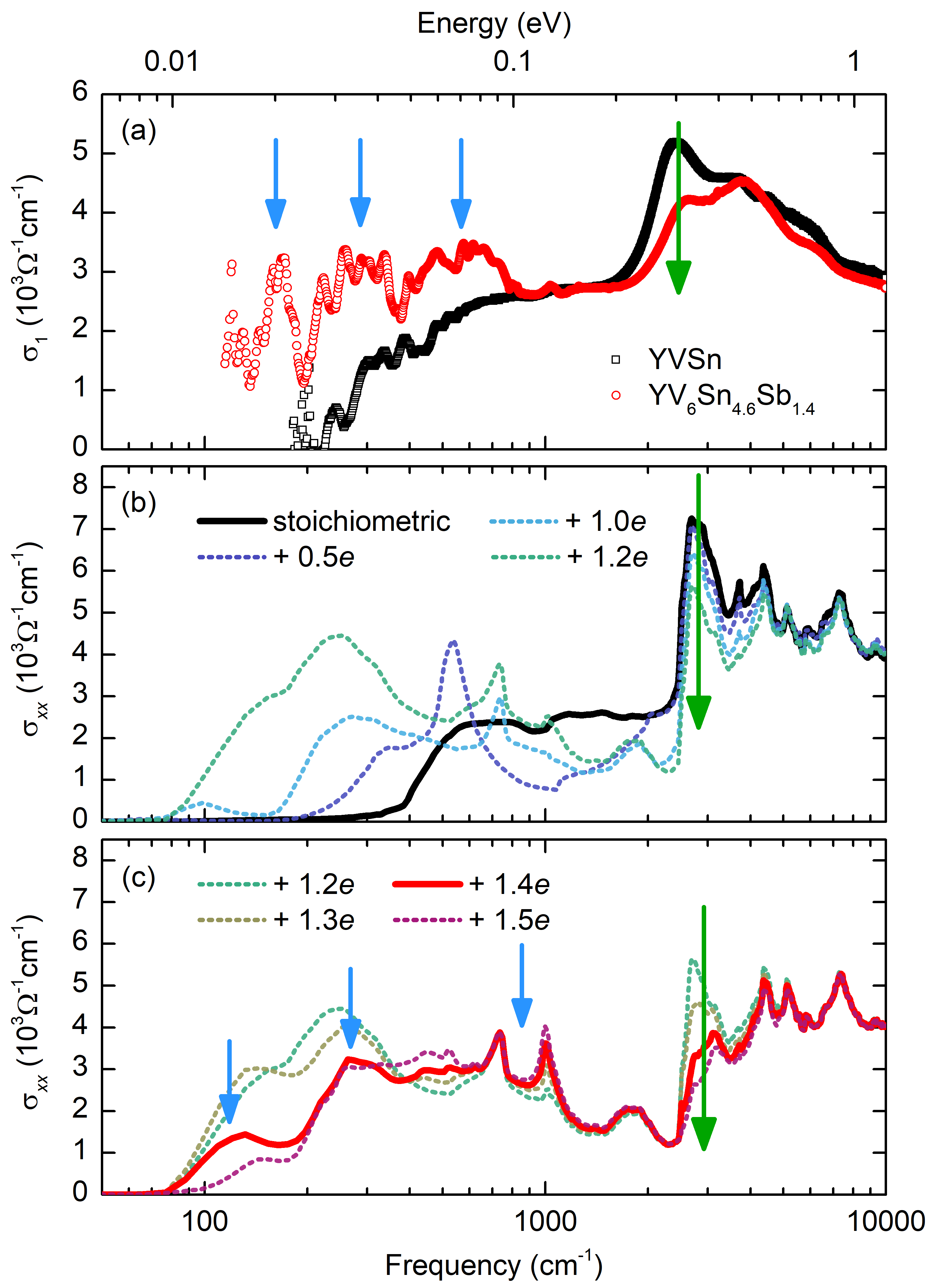}
	\caption{Interband contribution to the optical conductivity. (a) Experimental interband optical conductivity obtained by subtracting the modeled intraband contributions (Drude + localization peaks) and the phonon mode from the data. Characteristic modifications observed in the doped sample are highlighted by blue and green arrows. (b)(c) DFT-derived interband absorption for the stoichiometric case as well as for electron-doped calculations. The experimentally observed interband transitions in the doped case are well reproduced by adding 1.4 electrons per formula unit [red curve in panel (c)]. }		
	\label{doping_OC}
\end{figure}

We next examine the interband transitions in greater detail to identify changes in the optical spectra associated with doping-induced modifications of the electronic structure. Fig.~\ref{doping_OC}(a) shows the experimental interband optical conductivity, obtained by subtracting the intraband contributions (Drude and localization peaks) and, for pristine \YVS, the phonon mode from the spectra at 10~K. Upon doping, a clear reduction in spectral weight is observed in the mid-infrared range (green arrow). At the same time, three distinct absorption peaks emerge in the far-infrared spectrum of \doped\ (blue arrows), which are absent in pristine \YVS.

As indicated by the solid black line in Fig.~\ref{doping_OC}(b), the experimentally observed interband optical conductivity of pristine \YVS\ is well reproduced by our DFT calculations. The small discrepancies in the energies of the absorption features can be reduced by rescaling the frequency axis of the DFT-derived optical conductivity by a factor of $1/1.3$ (not shown in the figure), corresponding to a moderate renormalization of the band energies. For comparison, substantially larger renormalization factors have been reported for Mn-based 166 compounds~\cite{Gu2022, Wenzel2022}. This suggests that electronic correlations have a relatively weak effect on the band dispersion in \YVS, consistent with previous reports on V-based 166 kagome metals in general~\cite{Yang2024, Kim2023, Hu2023}.

To investigate the effects of electron doping on the interband optical conductivity, we carry out DFT calculations with progressively increasing numbers of electrons, thereby inducing a rigid shift of the Fermi level. The results are summarized in Fig.~\ref{doping_OC}(b) and (c) for selected doping levels. A systematic reduction of the spectral weight around 0.3~eV is observed with increasing electron doping (green arrow), while the low-energy interband transitions are much more sensitive to the position of the Fermi level. In particular, the low-energy spectral weight below 0.1~eV gradually increases up to electron dopings of 1.3 electrons/f.u., followed by a slight decrease at the lowest frequencies for even higher doping levels. The stoichiometry of the doped sample corresponds to a doping of 1.4 electrons/f.u. Indeed, this calculation reproduces the experimentally observed interband transitions remarkably well as highlighted by the corresponding arrows in panels~(a) and (c) of Fig.~\ref{doping_OC}. This agreement indicates that the main effects of Sb doping can be captured by a rigid shift of the Fermi level and, consequently, that the elemental substitution does not substantially modify the underlying electronic band dispersion.

\begin{figure*}
	\centering
	\includegraphics[width=2\columnwidth]{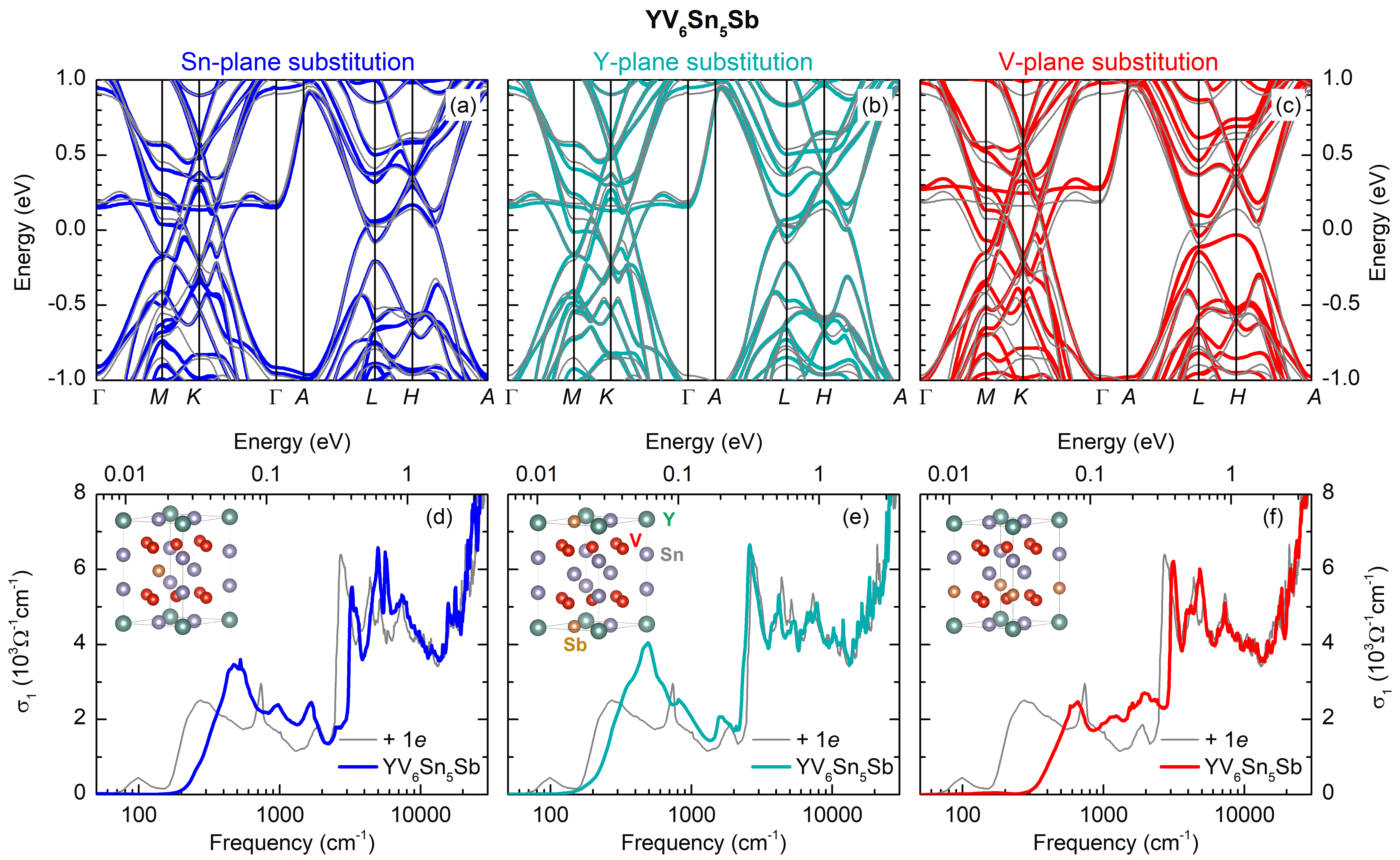}
	\caption{Electron-doped band structure with +1 electron/f.u. of \YVS\ (grey lines), together with the calculated band structures of YV$_6$Sn$_5$Sb, where one Sn atom per unit cell is substituted by Sb at the central Sn site~(a), the Y-layer Sn site~(b), or in the in-plane Sn site~(c). The corresponding optical conductivities are shown in panels~(d-f). The insets illustrate the respective substitution sites within the crystal structure.}		
	\label{DFT2}
\end{figure*}

The unit cell of \YVS\ contains three crystallographically inequivalent Sn sites, which occupy distinct layers along the $c$ axis. For clarity, we refer to these according to their structural environment as the central Sn layer (Sn2), the Y-layer Sn site (Sn3), and the in-plane Sn site (Sn1), with the latter being slightly displaced along the $c$ axis from the V kagome planes. We next perform DFT calculations for YV$_6$Sn$_5$Sb, in which one Sn atom per unit cell is replaced by Sb, corresponding to an increase of the electron count by 1 electron/f.u. 

Substitution at the central Sn site leaves the band dispersion largely intact, as shown in Fig.~\ref{DFT2}(a), and increases the low-energy spectral weight [Fig.~\ref{DFT2}(d)]. It does, however, shift the characteristic mid-infrared absorption at 0.3~eV to higher energies, which is not observed experimentally. For the Y-layer Sn site, the mid-infrared absorption remains essentially unchanged relative to the rigid-band shift expected for an increase of 1 electron/f.u., while the low-energy spectral weight is likewise enhanced [Fig.~\ref{DFT2}(e)]. The overall band dispersion also retains the characteristic features of pristine \YVS\ [Fig.~\ref{DFT2}(b)]. Substitution at the in-plane Sn site produces a markedly different response: the electronic structure deviates more strongly from a simple rigid shift of the Fermi level, as shown in Fig.~\ref{DFT2}(c), with the substitution modifying large parts of the band dispersion. This is also reflected in the optical conductivity, which exhibits reduced low-energy spectral weight and a shift of the mid-infrared absorption to higher energies; neither of which is observed experimentally.

While the electronic structure of V-based 166 compounds in the vicinity of the Fermi level is predominantly governed by V $3d$ states, the Sn1 $p$ states associated with the in-plane V kagome layer provide a significant contribution to the electronic states along $A$--$L$--$H$--$A$~\cite{Tsirlin2025, Cao2023, Yang2024, Shtefiienko2025}. This region of the band structure is indeed strongly affected by Sb substitution at the in-plane Sn1 site. The good agreement between our experimental results and the rigid-band-shift picture therefore indicates that the dispersions of the $p$-derived bands are not significantly modified by the experimentally relevant Sb substitution. This suggests that Sb substitution predominantly affects the out-of-plane Sn3 sites associated with the Y layer.
 
In contrast, doping the $p$-element site in the $A$V$_3$Sb$_5$ family can significantly modify the electronic properties of these compounds, often producing effects resembling those of hydrostatic pressure. This indicates substantial changes in the band dispersion, also including the vanadium-derived kagome bands~\cite{Salinas2026, Oey2022, Oey2022a, Liu2022, Salinas2023}. Such behavior is not unexpected given the stronger hybridization between V $3d$ and Sb $5p$ states in these systems~\cite{Tsirlin2025}.

\subsection{Band-Resolved Optical Conductivity}
\begin{figure*}
	\centering
	\includegraphics[width=2\columnwidth]{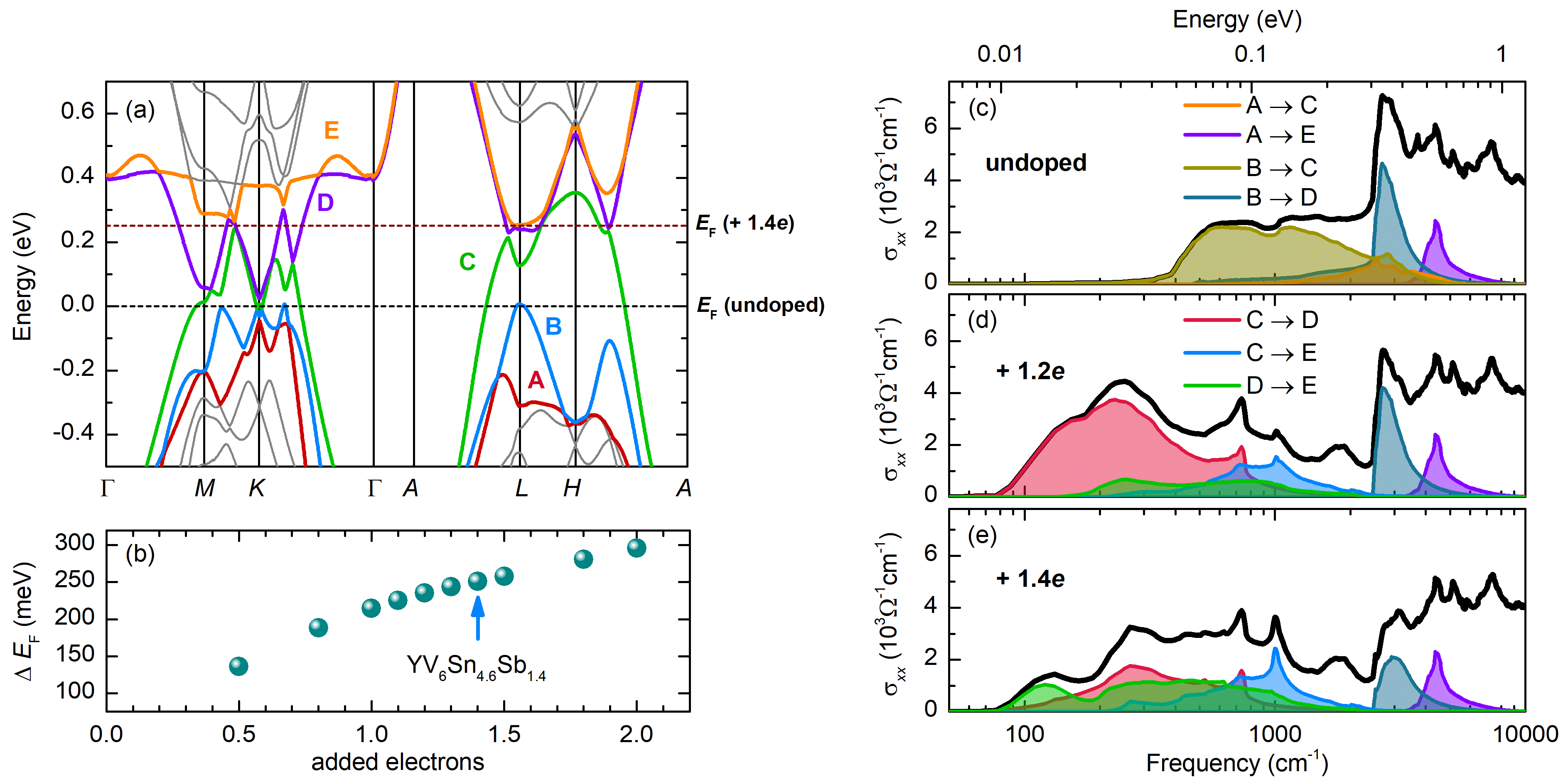}
	\caption{(a)~Band structure of \YVS. The Fermi level in the case of electron-doped calculations with +1.4 electrons/f.u. is marked by the brown dotted line. (b)~Fermi level shift as a function of number of added electrons in the calculations. (c)~Band-resolved optical conductivity in the undoped case, following the labels of the bands in panel~(a). Band-resolved optical conductivity for doped calculations with +1.2 electrons/f.u.~(d) and +1.4 electrons/f.u.~(e), with the latter showing the best agreement with the experimentally observed interband optical conductivity.}		
	\label{DFT}
\end{figure*}

To understand the evolution of the interband transitions and relate their changes to the position of the Fermi level, we carried out band-resolved optical conductivity calculations, which allow us to assign the observed interband absorption features to specific bands. In the undoped case, the Fermi level lies in close proximity to a band saddle point of band C at the $M$ point, as well as to a manifold of linearly crossing bands at the $K$ point [see Fig.~\ref{DFT}(a)]. As shown in panel~(c), the step-like onset of the interband transitions originates predominantly from transitions between bands B and C in the vicinity of $M$--$K$, where these bands form multiple Dirac-cone-like features. With increasing electron doping, these transitions become progressively suppressed as the Fermi level shifts away from band B. At the same time, spectral weight is transferred to lower energies, as low as $\sim$100~\cm, which can be attributed to transitions between bands C and D [see Fig.~\ref{DFT}(d)].

The experimentally observed suppression of spectral weight around 0.3~eV can be understood as the result of multiple contributions. First, transitions from B $\rightarrow$ C and A $\rightarrow$ C contribute to the background in this energy range [see Fig.~\ref{DFT}(c)]. Upon doping, and consequently moving the Fermi level away from the bands A and B, these transitions become suppressed. Second, the sharp absorption peak centered at 0.3~eV, which is mainly associated with B $\rightarrow$ D transitions near the $L$ point, is strongly suppressed for dopings above 1.3~electrons/f.u., when the Fermi level shifts above band D. At the same time, as the Fermi level approaches band E, low-energy transitions involving bands D and E become dominant, as seen in the band-resolved optical conductivity for 1.4~electrons/f.u. in panel~(e) of Fig.~\ref{DFT}.

Importantly, our band-resolved calculations demonstrate that the relatively large Fermi-level shifts enabled by Sn-site doping in \YVS\ allow the Fermi level to be brought into close proximity to various characteristic band features, such as multiple Van Hove singularities and the flat band E along $\Gamma$--$M$--$K$--$\Gamma$. This broad tunability of the Fermi level, while leaving the underlying band dispersion largely unaffected, enables a systematic study of band-filling-dependent electronic properties in a nonmagnetic kagome metal.

\section{Conclusions}
We have investigated the effects of Sn-site substitution on the electronic properties of nonmagnetic YV$_6$Sn$_{6-x}$Sb$_x$ using dc transport, infrared spectroscopy, and DFT calculations. Our results reveal increased metallicity in \doped, as evidenced by the lower dc resistivity and enhanced Drude contribution to the optical conductivity. The observed modifications in the interband absorption can be well described within band theory by a rigid shift of the Fermi level. Band-resolved DFT calculations identify the origin of these interband transitions, demonstrating that relatively large Fermi-level shifts can be achieved through Sb doping. This provides access to multiple characteristic features of the kagome band structure, including Van Hove singularities and flat band regions and thus, enables systematic investigations of band-filling-dependent electronic properties of nonmagnetic kagome metals.

\section{Acknowledgments}
Authors are grateful to Gabriele Untereiner for helping with the sample preparation and acknowledge the fruitful discussions with Savita Priya. M.W. is supported by IQST Stuttgart/Ulm via a project funded by Carl Zeiss foundation. S.D.W. gratefully acknowledges support via the UC Santa Barbara NSF Quantum Foundry funded via the Q-AMASE-i program under Award No. DMR-1906325. The work has been supported by the Deutsche Forschungsgemeinschaft (DFG) via DR228/51-3, and UY63/5-1.

\bibliography{YV6Sn6}

\end{document}